\documentstyle[11pt]{article}
\begin{document}
\vspace*{-2cm}
\noindent
\hspace*{11cm}
UG--FT--72/97 \\
\hspace*{11cm}
hep-ph/9703461 \\
\hspace*{11cm}
April 1997 \\
\begin{center}
\begin{large}
{\bf Measure of the size of CP violation in extended models \\}
\end{large}
~\\
J. A. Aguilar-Saavedra \\
{\it Departamento de F\'{\i}sica Te\'{o}rica y del Cosmos \\
Universidad de Granada \\
18071 Granada, Spain}
\end{center}
\begin{abstract}
In this letter we introduce a possible measure of the size of CP violation in
the Standard Model and its extensions, based on quantities invariant under the
change of weak quark basis. We also introduce a measure of the ``average size''
of CP violation in a model, which can be used to compare the size of CP
violation in models involving extra sequential or vector-like quarks, or
left-right symmetry.
\end{abstract}
\hspace{0.8cm}
\vspace*{0.5cm}

The definition of ``Maximal CP Violation'' and the related problem of finding 
an adequate measure of the size of CP violation in the Standard Model (SM)
has been object of interest
\cite{papiro1,papiro2,papiro3,papiro3b,papiro4} since
experiment \cite{papiro5} revealed that the phase appearing in the 
Cabibbo-Kobayashi-Maskawa (CKM) matrix \cite{papiro6} had to be large to 
explain the observed CP violation in the $\mathrm K^0$-$\mathrm \bar K^0$ 
system. As shown by Wolfenstein \cite{papiro1} any definition of maximal CP 
violation based on maximizing the phase appearing in the CKM matrix does not 
make sense because it depends on the 
parametrization of the CKM matrix itself. A phase 
of $\pi/2$ in one parametrization does not correspond to a phase of $\pi/2$
in another parametrization. Gronau and Schechter \cite{papiro2} used the 
construction of the CKM matrix as a product of three 
$2 \times 2$ unitary matrices 
({\em i. e.} the Murnaghan construction) to find a certain combination of 
phases (the ``invariant phase'') that remained invariant under most
rephasings
of the quark fields. However, this invariant phase depends on the adoption of
the Murnaghan construction as well as on the order in which the $2 \times 2$
matrices are multiplied. The formulation in terms of rephasing invariants of
the CKM matrix was carried out independently by Jarlskog 
\cite{papiro3,papiro3b} and Dunietz, 
Greenberg and Wu \cite{papiro4} leading to 
two different definitions of measures. 
In Refs.
\cite{papiro3,papiro3b} the condition of maximality is that some of the
quantities $a_{\mathrm CP}=2\, {\mathrm Im}\,(\alpha \beta^*)/(|\alpha|^2+
|\beta|^2)$, with either $\alpha=V_{ij} V_{kl}$, $\beta=V_{kj} V_{il}$ or
$\alpha=V_{ij} V_{il}^*$, $\beta=V_{kj} V_{kl}^*$ or $\alpha=V_{ij} 
V_{kj}^*$, $\beta=V_{il} V_{kl}^*$ are maximal, $V_{ij}$ being the $ij$
element of the CKM matrix. Note that the numerators of all the
quantities $a_{\mathrm CP}$ are equal up to an overall sign by the unitarity
of the $3 \times 3$ CKM matrix but the normalization differs.
In Ref. \cite{papiro4} CP violation is
said to be maximal when the products ${\mathrm Im}\; V_{ij} V_{kj}^* V_{kl}
V_{il}^*$ acquire its
maximum absolute value. This occurs when there is maximum
mixing in the CKM matrix: the modulus
of all the matrix elements is $1/\sqrt 3$
and the phase of $V_{ud} V_{cd}^* V_{cs} V_{us}^*$ is $2 \pi/3$.
Nevertheless as pointed out by Botella and Chau \cite{papiro6b} this simple
and elegant definition cannot be generalized to a higher number of quark
generations because unitarity of a $N \times N$ CKM matrix does not imply
that all the quantities ${\mathrm Im}\; V_{ij} V_{kj}^* V_{kl} V_{il}^*$
are equal in modulus, and one has the case that when maximizing two of them a
third one becomes zero. In this letter we follow the work of Refs. 
\cite{papiro3,papiro3b,papiro4} although 
our definitions will be based on quantities 
invariant not only under rephasings of 
the CKM matrix but under arbitrary quark 
basis transformations. 

The formulation of CP violation in terms 
of quantities invariant under a change 
of weak quark basis is complementary to the usual formulation in terms of CKM
phases, and seems more adequate to define a measure of CP violation.
In Ref. \cite{papiro3b} Jarlskog 
 realized that, if properly normalized, the quantity ${\mathrm 
Im}\det [M_u M_u^\dagger,M_d M_d^\dagger]$,
invariant under weak quark basis 
transformations, can be looked as a measure of CP violation in the SM. 
($M_{u,d}$ are the up and down quark mass matrices respectively.)
All CP violation 
observables depend in a perhaps complicated way on this quantity, which then 
gives us a measure of CP violation in the SM. 
However, the normalization is not 
uniquely determined and can lead to different measures.

To extend this formulation to a more general model, 
let $\{M_i\}$ be the set of 
quark mass matrices of the model, together with 
their transpose, conjugate and hermitian 
conjugate. These matrices are such that under a change of weak quark basis
$M_i \rightarrow U_i^\dagger M_i U'_i$ with $U_i$, $U'_i$ unitary matrices.
For instance, in the SM it is enough for 
our purposes to choose $\{M_i\}=\{M_u,M_d,M_u^\dagger,M_d^\dagger\}$. We can 
construct a quantity invariant under a 
change of weak quark basis by taking the 
trace of a product of $n$ matrices, ${\mathrm tr}\;M_1 \cdots M_n$, which
can be repeated, chosen in such a way
that $U_{i+1}=U'_i$, $\forall i$ (with 
$U_{n+1} \equiv U_1$). The number $n$ of matrices may not be arbitrary in 
specific models. Then CP conservation (for the detailed proof
see Ref. \cite{papiro7}) implies that the imaginary 
part of this trace, which we will call an ``invariant'' from now on, is real:
\begin{equation}
I \equiv {\mathrm Im\;tr}\; M_1 \cdots M_n = 0\,.
\label{ec:1}
\end{equation}
A set $\{ I \}$ of invariants can be constructed
in this way. The vanishing of 
all these invariants is a necessary 
condition for CP conservation. Furthermore, 
in specific models sets of invariants can 
be found with the property that their 
vanishing is also a sufficient condition for CP conservation 
\cite{papiro3b,papiro8,papiro9,papiro11}. For instance, in the SM the condition
${\mathrm Im\;tr}\; H_u^2 H_d H_u H_d^2=0$ with $H_{x}=M_{x} 
M_{x}^\dagger$ is sufficient for CP conservation \cite{papiro8} (note that in
the SM ${\mathrm Im} \det [H_u,H_d]=2\,
{\mathrm Im\;tr}\;H_u^2 H_d H_u H_d^2$ 
). In the SM with an arbitrary number of generations of quarks all the 
invariants are constructed with products of the hermitian matrices $H_{u,d}$.
Any CP violation observable must depend, 
perhaps in a complicated way, on these 
invariants. Generalizing the results in Refs. \cite{papiro3,papiro3b}, we argue
that these invariants properly normalized
(``reduced invariants'') give measures of CP violation.

As pointed above, the normalization is not unique. Nevertheless,
by imposing some 
properties on the reduced invariants we
can determine a plausible form. Let us 
consider the invariants
\begin{eqnarray}
I_1 & = & {\mathrm Im\;tr}\; H_u^2 H_d H_u H_d^2\,, \nonumber \\
I_2 & = & {\mathrm Im\;tr}\; H_u^3 H_d H_u H_d^2\,.
\label{ec:2}
\end{eqnarray}
In the SM, all the invariants are proportional to $I_1$, which is the lowest 
order nontrivial one. Explicit calculation shows that 
$I_2=(m_u^2+m_c^2+m_t^2)\,I_1$, with $m_i$ the mass of the quark $i$. If we 
require that the reduced invariants $\tilde I_1$, $\tilde I_2$ are equal, the
relative normalization must be $(m_u^2+m_c^2+m_t^2)={\mathrm tr}\;H_u$. The
same can of course be done with $I_3={\mathrm Im\;tr}\; H_u^2 H_d H_u H_d^3$.
This suggests that the adequate normalization for an invariant $I$ with $n_u$
powers of $H_u$ and $n_d$ powers of $H_d$ should be $\tilde I=I/({\mathrm
tr}\;H_u)^{n_u} ({\mathrm tr}\;H_d)^{n_d}$. 
This generalizes to other models as 
follows: for an invariant $I={\mathrm Im\;tr}\;M_1 \cdots M_n$ the reduced
invariant $\tilde I$ is
\begin{equation}
\tilde I=\frac{ {\mathrm Im\;tr}\;M_1 \cdots M_n }{ ({\mathrm tr}\;M_1 
M_1^\dagger)^{1/2} \cdots ({\mathrm tr}\;M_n M_n^\dagger)^{1/2}}\,.
\label{ec:3}
\end{equation}
This normalization has the remarkable property that $-1 \leq \tilde I \leq 1$
for any reduced invariant $\tilde I$ 
and arbitrary matrices $M_1,\dots,M_n$, as 
holds for the CP asymmetries $a_{\mathrm CP}$ discussed in Refs. 
\cite{papiro3,papiro3b}. This property can be easily shown using the Schwarz
inequality for the scalar product of matrices $\langle A,B \rangle
={\mathrm tr}\;AB^\dagger$ 
which in this case reads $({\mathrm tr}\;AB^\dagger)^2 \leq {\mathrm 
tr}\;AA^\dagger\; {\mathrm tr}\;BB^\dagger$.

A deeper reason to choose this normalization is that for a measure of CP 
violation we want to compare the CP violating invariant $I={\mathrm 
Im\;tr}\;M_1 \cdots M_n$ with the ``size'' 
(in some sense) of the mass matrices 
involved in its definition. The bare quantity $I$ cannot give this measure
because with a rescaling $M_i \rightarrow \lambda M_i$ it transforms as $I
\rightarrow \lambda^n I$ (because it has dimensions $[I]=m^n$) and
in absence of anomalies, physics should not change with such a scale
transformation. Thus we have to divide $I$ by a quantity ${\mathcal N}$ with
dimensions $[{\mathcal N}]=m^n$, invariant under a change of weak quark basis
and which in some sense measures the ``size'' of the mass matrices
$M_1,\dots,M_n$. The only nonsingular 
invariant measure of the size of a matrix 
$A$ is its norm $||A||=({\mathrm tr}\;A A^\dagger)^{1/2}$
(the other invariants of $A$, for instance its determinant,
can be zero without $A$ being identically zero and can lead to 
singularities in the definition). The simplest choice is then Eq. 
(\ref{ec:3}).

It is apparent that there is no set of parameters (masses, mixings, phases) 
that maximizes all the quantities $\tilde I$.
So, any definition of maximal CP 
violation based on reduced invariants 
will have the arbitrariness in the choice 
of the reduced invariants that are maximized. The most obvious
possibility is to 
choose the lowest order (lowest number of mass matrices in the numerator) 
reduced invariant, because in general higher order reduced invariants involve
higher powers of the mixing angles and therefore have smaller maximum values
than the lowest order ones. Having this in mind, we will find the maximum
value of the lowest order (nontrivial) reduced invariants for some models.

As we have seen before, $I_1$ in Eq. (\ref{ec:2}) is the lowest order
nontrivial invariant in the SM. Written as a function of physical parameters,
\begin{eqnarray}
I_1 & = & - (m_c^2-m_u^2)(m_t^2-m_u^2)(m_t^2-m_c^2)
(m_s^2-m_d^2)(m_b^2-m_d^2)(m_b^2-m_s^2)  \nonumber \\
& & \, \times \, {\mathrm Im}\;V_{ud} V_{cd}^* V_{cs} V_{us}^*\,.
\label{ec:3b}
\end{eqnarray}
and the reduced invariant $\tilde I_1$
decouples into a factor involving up quark
masses only, another one with down quark masses and a third one involving CKM
matrix elements:
\begin{eqnarray}
\tilde I_1 & = & - \frac{(m_c^2-m_u^2)(m_t^2-m_u^2)(m_t^2-m_c^2)}
{(m_u^2+m_c^2+m_t^2)^3} \times \frac{(m_s^2-m_d^2)(m_b^2-m_d^2)(m_b^2-m_s^2)}
{(m_d^2+m_s^2+m_b^2)^3} \nonumber \\
& & \, \times \, {\mathrm Im}\;V_{ud} V_{cd}^* V_{cs} V_{us}^*\,.
\label{ec:4}
\end{eqnarray}
The maximum value of the third factor in Eq. (\ref{ec:4}) was found in Ref.
\cite{papiro4} to be $1/6 \sqrt 3$. It is straightforward to show that the
maximum value of the up (down) factor occurs when one mass is zero and the
ratio of the other two is $(1+\sqrt 3)/\sqrt 2$, and is (surprisingly) also 
$1/6 \sqrt 3$. Then, the maximum value of $\tilde I_1$ is $(1/6 \sqrt 3)^3$.

This result can be compared with the hypothetical case that the top quark did
not exist, {\em i. e.} the bottom quark was a vector-like  
${\mathrm SU}(2)_L$ singlet.
In this model, the up quark mass matrix $M_u$ has dimensions $2 \times 2$ and
the $3 \times 3$ down quark mass matrix ${\mathcal M}_d$ divides in two
submatrices
\begin{equation}
{\mathcal M}_d=\left( \begin{array}{c} M_d \\ m_d \end{array} \right )\,.
\label{ec:5}
\end{equation}
$M_d$ is a $2 \times 3$ matrix connecting the
 two left-handed ${\mathrm SU}(2)_L$
doublets with the three right-handed
${\mathrm SU}(2)_L$ singlets; $m_d$ is a
$1 \times 3$ matrix connecting the left-handed
${\mathrm SU}(2)_L$ singlet with the
three right-handed ${\mathrm SU}(2)_L$ singlets.
(For a detailed description of the
models with quark singlets see Ref. \cite{papiro10}).
The lowest order nontrivial
invariant of this model is \cite{papiro11}
\begin{eqnarray}
I'_1 & = & {\mathrm Im\;tr}\;H_u H_d h_d h_d^\dagger \nonumber \\
& = & -(m_c^2-m_u^2)(m_s^2-m_d^2)(m_b^2-m_d^2)(m_b^2-m_s^2)\;
{\mathrm Im}\;V_{ud} V_{cd}^* V_{cs} V_{us}^*\,,
\label{ec:6}
\end{eqnarray}
with $H_x=M_x M_x^\dagger$, $h_d=M_d m_d^\dagger$. 
It has the same dependence in
down quark masses and CKM matrix elements than $I_1$ in Eq. (\ref{ec:3b}). 
The corresponding reduced invariant
is $\tilde I'_1=I'_1/({\mathrm tr}\;H_u)
({\mathrm tr}\;H_d)^2 ({\mathrm tr}\;m_d
m_d^\dagger)$, and its maximum value is
$1/16$, roughly 70 times larger than the
maximum value of $\tilde I_1$ in the SM. It is worthwhile to note that a
naive analysis involving only the rephasing invariant CKM factor in Eqs.
(\ref{ec:3b},\ref{ec:6}) yields in both
models a maximum value of $1/6 \sqrt 3$. This
does not reflect the fact that in the latter
model the effects of CP violation can be
larger than in the SM. The formulation
in terms of invariants is more adequate because
although in both models the CKM matrix
is a $3 \times 3$ unitary matrix (in the latter
case only the upper $2 \times 3$
submatrix is observable) with one CP violating phase,
the invariants are different in each
case and the observables depend on the masses and
CKM matrix elements through the invariants. 

These models with small CP violation effects can be compared with models with
left-right symmetry, where CP violation can
be very important. In these models, even
with one generation of quarks CP violation
is possible. The lowest order nontrivial
invariant is \cite{papiro12}
\begin{equation}
I''_1={\mathrm Im\;tr}\;M_u M_d^\dagger\,.
\label{ec:7}
\end{equation}
In the one generation case, $M_u$ and $M_d$
are complex numbers $m_u$ and $m_d$ respectively, and
the reduced invariant is 
$\tilde I''_1={\mathrm Im}\;(m_u m_d^*)/|m_u|\,|m_d|$,
which has a maximum value of 1 when
the relative phase between $m_u$ and $m_d$ is
$\pi/2$. In the three generation case
$\tilde I''_1$ also reaches the value 1 when the
up and down mass matrices are equal up to an overall factor.

We can go a step further in our analysis and try to define quantities that 
characterize the {\em average} size of CP violation in a specific model, 
independently of the {\em actual} size 
({\em i. e.} when one inserts the known 
values of masses, mixings, etc.) of CP violation. 
The significance of this will be clear later.
For this purpose we define a vector space 
${\mathcal I}$ whose elements $g$ are 
the reduced invariants of a given model, which form a basis in this 
space, plus real linear combinations of them. We can regard the elements of 
${\mathcal I}$ as functions of $k$ real 
variables, with $k$ twice the total 
number of (complex) entries in the quark mass matrices of the model. (For 
instance, in the SM $k=2 \times (3 \times 3 + 3 \times 3)=36$.) 
We consider the 
vector subspace ${\mathcal I}_0 \subset {\mathcal I}$ consisting of the 
elements of ${\mathcal I}$ which vanish identically for all values of its $k$
variables (the reduced invariants $\tilde I \propto {\mathrm Im\;tr}\;M$ with
$M$ hermitian are some of the elements of
${\mathcal I}_0$). Then we construct 
the vector space ${\mathcal I}'={\mathcal I}/{\mathcal I}_0$. In this space, 
but not in ${\mathcal I}$, we can define a norm
\begin{equation}
||g||=\left( \frac{1}{V(S)} \int_S |g(x)|^2 dx \right)^\frac{1}{2}\,,
\label{ec:8}
\end{equation}
where $x$ denotes the $k$ variables of $g$, $S$ is a subset of
${\mathrm \bf R}^k$ 
which we will specify later and $V(S)$ is the volume of $S$. 
For the moment, we 
can assume $S$ to be a ball centered at the origin and of arbitrary 
radius $r$. With this definition, we see: (i) $||g|| \geq 0$; (ii) If 
$||g||=0$, $g(x)=0$ in $S$, but our definition 
of $S$ implies $g(x)=0$ $\forall x 
\in {\mathrm \bf R}^k$ and $g \in {\mathcal I}_0$; (iii) $|| \alpha g 
||=|\alpha|\,||g||$ with $\alpha$ a real number; and (iv) $||g_1+g_2|| \leq 
||g_1||+ ||g_2||$ as a direct consequence 
of the same property for functions of 
$k$ variables. Thus $|| \cdot ||$ defines a norm on ${\mathcal I}'$. Note
that the property of the reduced invariants $-1 \leq \tilde I \leq 1$ implies
that the integrand is bounded, so the integral is well defined. At this point
it is important to note that if we choose another normalization for the
invariants, we can define the vector spaces $\mathcal I$, ${\mathcal I}_0$,
${\mathcal I}'$ in the same way and the operation $|| \cdot ||$ is still a
norm although the modulus of the vectors differs. 

Before we give a significance to the norm we will specify the set $S$. This 
is best explained with an example. Let us consider the SM with quark mass 
matrices $M_u=(m_{ij})$, $M_d=(n_{ij})$. The elements of the set $S$ 
discussed above are points $x \in {\mathrm \bf R}^k$ with components
${\mathrm Re}\;m_{ij}$, ${\mathrm Im}\;m_{ij}$, ${\mathrm Re}\;n_{ij}$, 
${\mathrm Im}\;n_{ij}$, $i,j=1,2,3$ such that $\sum_{ij} {\mathrm Re}^2 m_{ij} +
{\mathrm Im}^2 m_{ij} + {\mathrm Re}^2 n_{ij} + {\mathrm Im}^2 n_{ij} 
\leq r^2 $. This set is not invariant under weak basis transformations, so it
is more convenient to define $S=\{x / 
\sum_{ij} {\mathrm Re}^2 m_{ij} + {\mathrm Im}^2 m_{ij} \leq r^2,\;\sum_{ij}
{\mathrm Re}^2 n_{ij} + {\mathrm Im}^2 n_{ij} \leq r^2 \}$ {\em i.e.} the set
$S$ is the direct product of two 18-dimensional balls centered at the origin.
The norm $||g||$ of a vector is independent
of the value of $r$ because $g$ is dimensionless.
In a general model, the set $S$ can be taken to be
the direct product of a ball of radius 1 centered at the origin containing
the parameters for each mass matrix. Then,
the significance of $||\tilde I||$ is: (i) Choose a point in the 
parameter space of the theory; (ii) Calculate the measure of CP violation in
this theory with these parameters, 
as given by $\tilde I$; (iii) Average over a 
symmetric set of points containing all possible values of the parameters. The
result is an ``average measure'' of CP violation in the model, as given by
$\tilde I$. The interest of these average measures constructed as norms of
reduced invariants is that they tell us
{\em a priori} and without knowledge of 
the values of the parameters in the theory whether CP violation effects are
expected to be large or small. Of course, the size of CP violation effects is
given by the actual parameters, but these average measures give us the global
behavior of the theory. Comparing the norms of the reduced invariants of two
theories we can state in which theory CP violation is more important. The 
numerical calculation of the norms of $I_1$, $I'_1$, $I''_1$ (the latter for
three generations of quarks) yields
\begin{eqnarray}
|| I_1 || & = & 1.63 \cdot 10^{-5} \,, \nonumber \\
|| I'_1 || & = & 2.51 \cdot 10^{-4} \,,\nonumber \\
|| I''_1 || & = & 0.4003 \,.
\label{ec:9}
\end{eqnarray}
The evaluation of the multidimensional integral of Eq. (\ref{ec:8}) to obtain
Eqs. (\ref{ec:9}) is a difficult task because the integrand is very
oscillatory (specially in the first case) and we integrate over a huge
number of dimensions (36, 26 and 36 respectively), so we have used a Monte
Carlo method and the errors are on the last decimal place. We see
that $|| I_1 || < || I'_1 ||$ and $|| I'_1 || < || I''_1 ||$ as expected from
the hierarchy of their maximum values. However, while in the model with
left-right symmetry we have $|| I''_1 ||/(I''_1)_{\mathrm max} \simeq
0.4$, in the other two models we have $|| I_1 ||/(I_1)_{\mathrm max} \simeq
1.8 \cdot 10^{-2}$, $|| I'_1 ||/(I'_1)_{\mathrm max} \simeq 4.0 \cdot
10^{-3}$ (so the two first integrals are strongly peaked and difficult to
calculate). This means that the {\em average} size of CP violation in the
SM and the model with vector-like bottom is much less than the maximum
allowed by the model, {\em i. e.} by the quark content and symmetries of
the Lagrangian. These results are expected to be general and not sensitive
to the choice of normalization. Within the SM, using the experimental values
of masses
and mixings we find $I_1 \leq 9.1 \cdot 10^{-13}$. Thus, in the SM not only
CP violation is much less than maximal, but is also much less than the
average.

\newpage
\noindent
{\Large \bf Acknowledgements}

\vspace{0.4cm}
\noindent
I wish to thank G. C. Branco and the Instituto Superior T\'{e}cnico
of Lisboa for their kind hospitality during the realization of this work. I am 
indebted to F. del Aguila for discussions.
I also thank J. I. Illana and M. Baillargeon for help on
Monte Carlo techniques. This work was
partially supported by CICYT under contract AEN94-0936, by the Junta
de Andaluc\'{\i}a and by the European Union under contract
CHRX-CT92-0004.

\end{document}